\RequirePackage{fix-cm}
\documentclass[final,twocolumn,nofootinbib,longbibliography]{svjour3}          
\usepackage{graphicx}
\usepackage[dvipsnames,table,xcdraw]{xcolor}
\usepackage{etoolbox}

\usepackage{comment}
\usepackage{multirow}
\usepackage[utf8]{inputenc}
\usepackage{amsmath}
\usepackage{hyperref}
\usepackage[english]{babel}
\usepackage{microtype}
\usepackage{ulem}
\usepackage{stfloats}
\usepackage{afterpage}
\hypersetup{colorlinks=true,
	linktocpage=true,
	breaklinks=true,
	bookmarksnumbered,
	bookmarksopen=true,
	bookmarksopenlevel=1,
	hypertexnames=true,
	pdfhighlight=/0,
	urlcolor=black!38!red,
	linkcolor=MidnightBlue!60!Black,
	citecolor=PineGreen!60!Black
}
\usepackage{url}
\usepackage[utf8]{inputenc}
\usepackage{physics}
\usepackage{tensor}
\usepackage{amssymb}
\usepackage{bbold}
\usepackage{orcidlink}
\usepackage{authblk}

\begin{document}

\title{Implementation and readout of maximally entangled two-qubit gates quantum circuits in a superconducting quantum processor}

\titlerunning{  }

\author{V. Stasino\orcidlink{0009-0009-4898-5834}$^{1}$ \and 
P. Mastrovito\orcidlink{0000-0002-7833-0398}$^{1,2}$ \and 
C. Cosenza\orcidlink{0009-0004-2426-4008}$^{1}$ \and 
A. Levochkina\orcidlink{0009-0002-2615-2507}$^{1,2}$ \and 
M. Esposito\orcidlink{0000-0002-7674-9195}$^{2}$ \and 
D. Montemurro\orcidlink{0000-0001-8944-0640}$^{1,2}$ \and 
G.P. Pepe\orcidlink{0000-0002-7302-4412}$^{1,2}$ \and 
A. Bruno\orcidlink{0000-0002-7821-5082}$^{3}$ \and 
F. Tafuri\orcidlink{0000-0003-0784-1454}$^{1}$ \and 
D. Massarotti\orcidlink{0000-0001-7495-362X}$^{4}$ \and 
H.G. Ahmad\orcidlink{0000-0003-2627-2496}$^{1,2}$}

\authorrunning{   }

\institute{
    \textsuperscript{1}Dipartimento di Fisica Ettore Pancini, Università degli Studi di Napoli Federico II, c/o Complesso Monte Sant’Angelo, via Cinthia, I-80126 Napoli, Italy \\
    \textsuperscript{2}CNR-SPIN, Complesso di Monte S. Angelo, via Cintia, 80126, Napoli, Italy \\
    \textsuperscript{3}QuantWare, Elektronicaweg 10, 2628 XG Delft, The Netherlands \\
    \textsuperscript{4}Dipartimento di Ingegneria Elettrica e delle Tecnologie dell'Informazione, Università degli Studi di Napoli Federico II, 80125, Napoli, Italy
}


\maketitle

\begin{abstract}

Besides noticeable challenges in implementing low-error single- and two-qubit quantum gates in superconducting quantum processors, the readout technique \textcolor{black}{and analysis are} a key factor in determining the efficiency and performance of quantum processors. Being able to efficiently implement quantum algorithms involving entangling gates and \textcolor{black}{asses} their output  is mandatory for quantum utility. In a transmon-based $5$-qubit superconducting quantum processor, we compared the performance of quantum circuits involving an increasing level of complexity, from single-qubit circuits to maximally entangled Bell circuits. This comparison highlighted the importance of the readout \textcolor{black}{analysis} and helped us optimize the protocol for more advanced quantum algorithms. Here we report the results obtained from \textcolor{black}{the analysis of} the outputs of quantum circuits using two readout \textcolor{black}{paradigms}, referred to as \textcolor{black}{"multiplied readout probabilities" and "conditional readout probabilities"}. The first method is suitable for single-qubit circuits, while the second is essential for accurately \textcolor{black}{interpreting} the outputs of circuits involving two-qubit gates.

\end{abstract}

\section{Introduction}
The macroscopic quantum nature of Josephson junctions (JJs) is the key for the development of sophisticated quantum and classical cryogenic circuits, able to perform several types of operations: parametric amplification~\cite{Mallet2009,Yaakobi2013,Schmitt2014,Walter2017,Esposito2021,Aumentando2020}, single-photon generation and detection~\cite{Dauler2014,Salvoni2022}, novel microwave readout and control with cryogenic electronics\textcolor{black}{~\cite{MIURA2012,Leonard2019,Wang2023,DiPalma2023,Howe2022,Castellanos-Beltran2023,VanDijk2020}}, protocols including cryogenic memories~\cite{Alam2023,Satariano2024} and quantum computation~\cite{arute2019,kim2023evidence,Fowler2012,Bravyi2022}.

Superconducting Quantum Processing Units (sQPU) provide encoding and manipulation of quantum information in single superconducting quantum bits and allow to implement up to thousands consecutive quantum logic operations (quantum gates)~\cite{kwon2021}, whose maximum number gets higher the longer are the coherence times of the system~\cite{Siddiqi2021,Wang2022,somoroff2023}. Most importantly, in sQPU, it is possible to experimentally implement the most straightforward evidences of quantum mechanics on a macroscopic scale: quantum superposition and entanglement generation~\cite{García-Álvarez2017,Mooney2019}.

Despite significant recent progress with Noisy Intermediate Scale Quantum (NISQ) computers~\cite{Morvan2024}, running efficient quantum algorithms on sQPUs for practical goals is still limited by the current hardware capabilities. Excluding leading commercial players like Google or IBM~\cite{Qubits2024}, most of the sQPUs available and reported in literature are composed of around ten qubits~\cite{Harris2010,Chen2022,Ronkko2024}, and the coupling with the external environment plays a fundamental role in quantum computing efficiency. Moreover, scaling the number of qubits in the same platform has been demonstrated to introduce loss of quantum information (state-leakage) and qubit-qubit spurious crosstalk, which is the main cause of errors, thus severely limiting the race toward quantum utility~\cite{preskill2018quantum}.

Therefore, a strong effort from the user is required to correct and mitigate errors that may occur in the sQPU to run efficient quantum algorithms. Most importantly, there is still a strong dependence between hardware and software: hardware capabilities must improve to meet the request for the implementation of application-specific quantum algorithms, and software must be coded and tested in such a way that knowledge of the hardware is taken into consideration.

In this work, we aim at providing insightful details on the experimental techniques used to implement and readout single- and two-qubit gates algorithms on a five-qubit sQPU based on superconducting transmon qubits. Most importantly, it is crucial to establish an experimental protocol to efficiently \textcolor{black}{interpret} the output of a quantum algorithm. We discuss the difficulties arising when passing from single-qubit gate algorithms to highly-entangled quantum states generation. We demonstrate that \textcolor{black}{a crucial role is played by the method used to interpret the output of a quantum circuit, specifically when dealing with} the most entangled quantum states, the Bell states~\cite{steffen2006,Majer2007,horodeki2009}, marking a key benchmark for the implementation of more sophisticated quantum algorithms.

\begin{figure*}[t]
    \begin{center}
\includegraphics[width=1\textwidth]{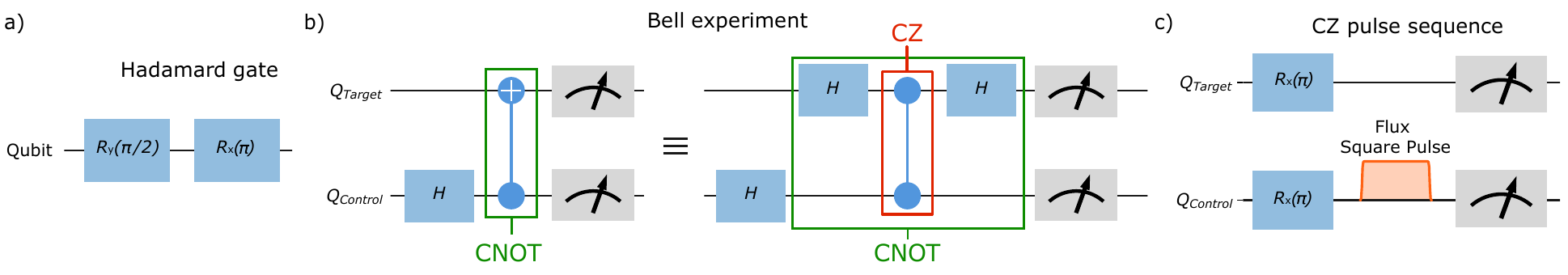}
    \end{center}
    \caption{In a) \textcolor{black}{equivalent decomposition of} a Hadamard gate. A \textit{$\textcolor{black}{R_y(\pi/2)}$} pulse is applied, followed by a \textit{$\textcolor{black}{R_x(\pi)}$}-pulse~\cite{qinspire}. In b), the pulse scheme of the Bell states circuit. A Hadamard gate is applied on the control qubit, followed by a \textit{CNOT} gate (green). The \textit{CNOT} gate can be decomposed in terms of Hadamard and \textit{CZ} gates. The pulse sequence is composed of a Hadamard applied on target, a \textit{CZ} (red), which is symmetric, and finally another Hadarmad on target. \textcolor{black}{In c) pulse sequence of \textit{CZ} experiment. An $R_x(\pi)$-pulse is applied both on the control and target qubits so that the state of the system is $\ket{11}$. Then, the control qubit is tuned with a flux pulse with variable amplitude and duration. Finally, a simultaneous measurement of the qubits is performed.}}
    \label{bell_circuit}
\end{figure*}

\section{Implementing single- and two-qubit gates: towards Bell state generation in NISQ superconducting devices}

A Bell pair is a maximally entangled 2-qubit quantum state~\cite{Nielsen2010}, specially useful for applications like: quantum computing and communication~\cite{steffen2006,Majer2007,sych2009complete,Nielsen_Chuang_2010,zaman2018counterfactual}, quantum teleportation~\cite{Kim2000,storz2023loophole}, generation of multipartite entangled states that allow to work with more than two qubits at a time for scalable quantum algorithms~\cite{wang2005quantum,horodeki2009,Mooney2019,ma2020manipulating,wang2022flying,cao2023generation}, superdense coding~\cite{harrow2004superdense,kang2016} and quantum criptography~\cite{boileau2004,wang2005quantum}.

There are four types of Bell states, which are the maximally entangled quantum states that can be implemented with two qubits. They represent the simplest example of entanglement. More specifically, the Bell states are:
\begin{equation}
\hspace{1.5cm}
    \ket{\Phi^+} = \frac{1}{\sqrt{2}} (\ket{0}\otimes\ket{0}+ \ket{1}\otimes\ket{1}),
    \label{00}
\end{equation}
\begin{equation}
\hspace{1.5cm}
    \ket{\Phi^-} = \frac{1}{\sqrt{2}} (\ket{0}\otimes\ket{0}- \ket{1}\otimes\ket{1}),
    \label{10}
\end{equation}

\begin{equation}
\hspace{1.5cm}
    \ket{\Psi^+} = \frac{1}{\sqrt{2}} (\ket{0}\otimes\ket{1}+ \ket{1}\otimes\ket{0}),
    \label{01}
\end{equation}
\begin{equation}
\hspace{1.5cm}
    \ket{\Psi^-} = \frac{1}{\sqrt{2}} (\ket{0}\otimes\ket{1}- \ket{1}\otimes\ket{0}).
    \label{11}
\end{equation}\\

Creating a Bell state typically involves single- and two-qubit quantum gates, like the Hadamard (\textit{H}) gate and the Controlled-NOT (\textit{CNOT}) gate (Fig.~\ref{bell_circuit}).
\textcolor{black}{Single-qubit gates can be expressed as rotation operators:
\begin{equation}
\hspace{2.7cm}
    R_{\alpha}(\theta) = e^{-i\frac{\theta}{2}\sigma_{\alpha}},
\end{equation}
where $\sigma_{\alpha}$, with $\alpha=\{x,y,z\}$, represents a Pauli matrix and $\theta$ is the rotation angle.
Specifically, a Hadamard gate is a single-qubit gate, which allows the generation of the superposition of the two basis states $\ket{0}$ and $\ket{1}$. It can be decomposed into native gates as~\cite{krantz2019}:
\begin{equation}
\begin{split}
    H = Ph_{\frac{\pi}{2}}R_y(\pi/2)R_z(\pi) = \hspace{0.8cm} \\
\hspace{1.2cm}
    =i\frac{1}{\sqrt{2}} \begin{bmatrix}
        1 & -1 \\
        1 & 1 
    \end{bmatrix}
    \begin{bmatrix}
        -i & 0 \\
        0 & i 
    \end{bmatrix} = \frac{1}{\sqrt{2}} \begin{bmatrix}
        1 & 1 \\
        1 & -1
    \end{bmatrix},
    \label{Hpulse}
\end{split}
\end{equation}
where $Ph_{\pi/2} = e^{i\frac{\pi}{2}} \textcolor{black}{\mathbb{1}}$ applies an overall phase $\pi/2$ to the qubit~\cite{krantz2019}.
In this work, we \textcolor{black}{used an equivalent decomposition for the Hadamard gate} more suitable for our instrumentation, detailed in App.~\ref{App1}. This circuit for the Hadamard gate involves a \textit{\textcolor{black}{$R_y(\pi/2)$}}, followed by an \textit{\textcolor{black}{$R_x(\pi)$}}~\cite{qinspire}, as shown in Fig.~\ref{bell_circuit} (a).}

The \textit{CNOT} gate is a two-qubit gate (Fig.~\ref{bell_circuit} (b))\textcolor{black}{,} referred to as conditional \textcolor{black}{gate}~\cite{krantz2019}. \textcolor{black}{Conditional gates} take two qubits as inputs: the first is called "control" and the second is called "target". The action of the target qubit strongly depends on the state of the control one. The \textit{CNOT} gate, for example, inverts the state of the target qubit when the control is excited, and it keeps it in the initial state if the control qubit is prepared in state $\ket{0}$. 

It is well-known that there are several possible decompositions of a \textit{CNOT} gate on a hardware level. The circuital design plays here a fundamental role: considering how the coupling between two or more qubits is engineered, as well as what kind of gate is possible to implement with the lowest error and in the fastest possible way, the \textit{CNOT} gate can be built by exploiting what are known as native two-qubit gates~\cite{krantz2019}. In sQPUs, iSWAP and Conditional-Z (\textit{CZ}) gates are the most common. 

In this work, we have used \textit{CZ} gates\textcolor{black}{~\cite{krantz2019,Rol2019}}. The \textit{CZ} is a unitary and symmetric gate, with the following matricial representation in the two-qubit computational basis:
\begin{equation}
\hspace{2.2cm}
   U_{CZ} = \begin{bmatrix}
    1 &0  &0  &0 \\
    0 &1  &0  &0\\
    0 &0  &1  &0\\
    0 &0 &0 &-1 \\
    \end{bmatrix}
    \label{CZmatrix}.
\end{equation} \\
It is possible to demonstrate that a \textit{CZ} gate allows to implement a \textit{CNOT} if combined with two Hadamard gates~\cite{krantz2019}\textcolor{black}{, as schematized in Fig.~\ref{bell_circuit} (b)}. The \textit{CNOT} unitary matrix reads as:
\begin{equation}
\hspace{1.5cm}
    U_{CNOT} = (\textcolor{black}{\mathbb{1}} \otimes H) U_{CZ} (\textcolor{black}{\mathbb{1}} \otimes H),
\end{equation}
since \textcolor{black}{$U_{CNOT} = \ket{0}\bra{0}\otimes \mathbb{1} + \ket{1}\bra{1}\otimes R_{x}(\pi)$, $U_{CZ} = \ket{0}\bra{0}\otimes \mathbb{1} + \ket{1}\bra{1}\otimes R_{z}(\pi)$ and $HR_{z}(\pi)H = R_{x}(\pi)$.}
In Fig.~\ref{bell_circuit} (b) it is shown the circuit diagram of the \textit{CNOT} in terms of a \textit{CZ}. 
\textcolor{black}{To perform a \textit{CZ} gate, it is necessary to excite both the qubits, so that the system is in $\ket{11}$~\cite{Negirneac2021}. The $\ket{11}$ state preparation is achieved by applying $R_x(\pi)$ rotations on both qubits simultaneously, as schematized in Fig.\ref{bell_circuit} (c). Then, it is required to set on resonance the two qubits. One possible way is to use flux-tunable transmon devices. }

In standard superconducting transmon \textcolor{black}{qubits}, which is also the circuital design employed in this work, Al/Al$O_x$/Al Josephson junctions are the main building blocks~\cite{krantz2019}. Here, the qubit frequency is related to the Josephson energy $E_J$, and therefore to the critical current of the JJ, and the charging energy $E_c$, which takes into account the total circuital capacitance of the device~\cite{koch2007}.
\textcolor{black}{In flux-tunable transmons, DC-SQUIDs (Superconducting Quantum Interference Device) are employed in place of a single JJ~\cite{koch2007}. As a matter of fact, in a DC-SQUID the Josephson energy can be tuned using an external flux $\phi$ as:
\begin{equation}
\hspace{2cm}
E_J(\phi)=E_J(0)\cos\left(\frac{\pi\phi}{\phi_0}\right),
\end{equation}
with $\phi_0$ the magnetic flux quantum~\cite{Barone1982,tafuri2019}. Therefore, also the qubit frequency ($\nu_{01} = \sqrt{8E_{J}E_{C}}-E_{C}$) can be tuned by changing $\phi$. Specifically, by engineering flux pulses one can implement a $R_z(\pi)$ gate on the target qubit when the control qubit is in the excited state, thus resulting in a swap of controlled excitation between the higher non-computational levels $\ket{11}$ and $\ket{20}$ of the two-qubit register.}
Considering standard values in transmons for the $E_J$ of few tens of gigahertz, and $E_c$ of the order of a few hundreds of megahertz, the typical range of qubit frequency is around $4-6$ GHz~\cite{Ahmad2023}. 

\textcolor{black}{The ability to implement low-error single-qubit gates is crucial for implementing two-qubit gates like the CZ. Indeed,} the implementation of XY single-qubit gates does not just rely on the frequency of the RF tone, which must resonate with the qubit frequency $\nu_{01}$, but requires to properly design fast drive pulses. While the frequency of the control signal allows to induce transitions between the two level states of a qubit, the amplitude and the duration of the pulses allow to set the rotation angles around the X and Y axes of the Bloch sphere~\cite{krantz2019}. At the same time, the phase of the signals allows to select rotations around the X and Y axes~\cite{krantz2019}. Finally, also the shape of the drive pulses plays a fundamental role. Provided that transmon qubits can be considered as weakly anharmonic quantum oscillators, where higher order energy levels are energetically close to the qubit frequency, state leakage outside the computational space is not excluded, as well as phase errors~\cite{reed2013,krantz2019}. In this case, one can change the shape of the pulse to efficiently suppress such spurious effects~\cite{Lucero2008,Motzoi2009,reed2013,Werninghaus2021,Babu2021}. Therefore, to implement single-qubit gates, such as the \textit{H} gate, it is required to calibrate all of these parameters. 

Finally, it is of paramount importance to provide a suitable way to infer the quantum state through readout \textcolor{black}{analysis}. In most transmon-based sQPUs, \textcolor{black}{readout} is achieved by capacitively coupling superconducting resonators to the qubits~\cite{koch2007}. It is well-known that by coupling a boson-like quantum field (a superconducting resonator is nothing else than a harmonic quantum oscillator) and a two-level quantum system, it is possible to make them exchange energy coherently depending on the coupling strength and the frequency detuning between the two systems~\cite{greentree2013fifty}. In transmon devices, it is preferred to work in the weak coupling regime, i.\,e. for $g/\Delta\ll1$, where $g$ is the coupling strength and $\Delta=\nu_{r}-\nu_{01}$ is the difference between the resonance frequencies of the qubit and the superconducting resonator~\cite{koch2007}. By circuitally engineering coupling strengths of the order of a few tens of megahertz, and qubit-resonator detunings of the order of the gigahertz, the weak coupling regime is established, and the readout resonator frequency is just perturbatively affected by any changes in the qubit state. Specifically, a dispersive shift \textcolor{black}{$\chi=g^2/\Delta$} arises~\cite{koch2007}, and it is possible to assess indirectly the state of the qubit by interrogating the response to an RF tone of the readout resonator, allowing the quantum non-demolition readout of the qubit state.

As follows, we provide details on the sQPU used in this work, as well as on the experimental methods and techniques used to implement Bell states, with a special focus on the readout procedure. Specifically, we here focus on the simultaneous dispersive readout of a Bell state implementation on the two-qubit computational basis, i.\,e. on the z-projection of a Bell state. We compare two statistical approaches for distinguishing between different output states, referred to as \textcolor{black}{”multiplied readout probabilities”} and \textcolor{black}{”conditional readout probabilities” paradigms}. We discuss that special care must be used when dealing with highly entangled output states to efficiently read the output of algorithms that employ Bell state preparation gates, and more generally two-qubit gates.

\section{Methods}
\subsection{Sample and experimental setup}
\label{Sec2}

\begin{figure}[t]
    \begin{center}
    \includegraphics[width=0.48\textwidth]{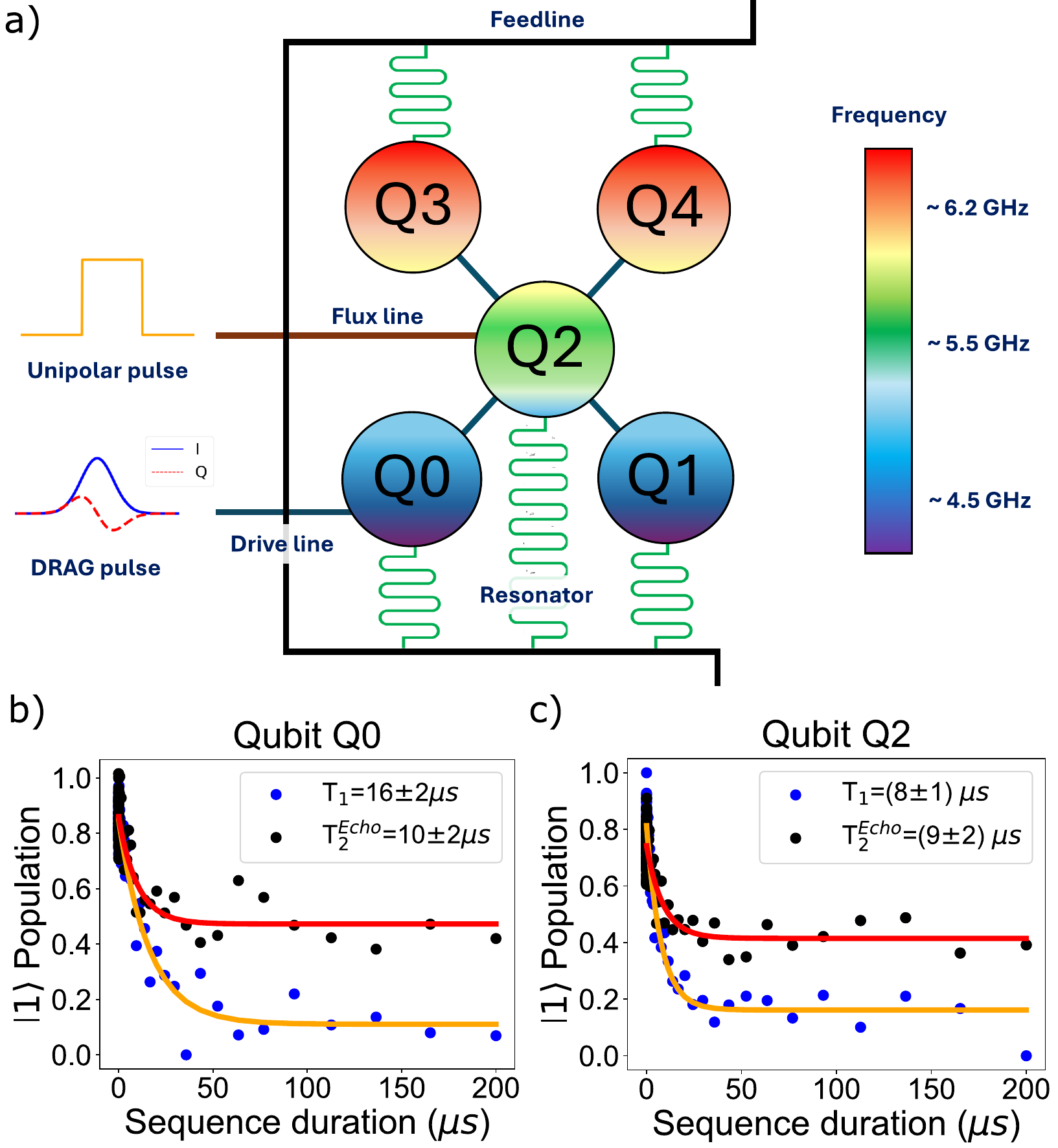}
    \end{center}
    \caption{Schematic layout of the QuantWare 5-qubit quantum processing unit Soprano. The chip consists of five coupled qubits with their readout resonators (green lines). Each qubit is provided with a drive line (blue line on Q0) and a flux line (brown line on Q2), drawn as an example only for Q0 and Q2, respectively. Each readout resonator is coupled to the common feedline (black line). The qubit-qubit coupling through Q2 is realized through high-frequency coupling resonator. On the left, the shapes of the pulses used for driving the qubits are shown. Specifically, to implement \textit{XY} pulses, DRAG (Derivative Reduction Adiabatic Gate) shaped pulses have been employed, while standard unipolar pulses are used for flux pulses. On the right, a schematic representation of the 5-qubit chip frequencies: Q0 and Q1 have the lowest resonance frequency (blue color shading) on the chip, Q3 and Q4 have the highest resonance frequency (orange color shading) and Q2 has an intermediate resonance frequency (green color shading). The color bar indicates the approximate values of the qubit frequencies.
    \textcolor{black}{In b) and c) $T_1$ and $T_2^{Echo}$ one-shot measurements for Q0 and Q2, respectively: the x-axis represents the sequence duration, while the y-axis the population of the excited state. The black dots and solid red line correspond to the measured values and the fit for $T_2^{Echo}$, respectively. Similarly, the blue dots and solid orange line correspond to the measured values and the fit for $T_1$. The legends show the results of $T_1$ and $T_2^{Echo}$ estimated from the fitting procedure.}}
    \label{chip5qubit}
\end{figure}

The sQPU (Soprano, manufactured by QuantWare) is composed of six transmon qubits: one isolated qubit for test, and 5 coupled qubits. More specifically, all the qubits are connected through a central qubit (Q2) by means of four high-frequency bus resonators~\cite{Majer2007,Sillanpää2007} in a star-like fashion. A schematic of the chip is shown in Fig.~\ref{chip5qubit} (a). Q0 and Q1 have the lowest resonance frequency on the chip, Q3 and Q4 have the highest resonance frequency, and Q2, i.\,e. the central qubit, has an intermediate resonance frequency. This design is suitable for advanced quantum error detection and correction~\cite{Kelly2014,versluis2017,Fowler2012}. All the qubits are flux-tunable, i.e., they include standard aluminum-based DC-SQUIDs. Each of the qubits has its own flux line for qubit frequency tunability, the drive line for control and the readout resonant cavity~\cite{koch2007,krantz2019}. The readout resonators (highlighted in green in Fig.~\ref{chip5qubit} (a)) are all coupled to a common feedline (black line in Fig.~\ref{chip5qubit} (a)). This configuration allows for multiplexing, i.e., to address multiple resonators with the single feedline by sending readout tones with different frequencies~\cite{Reed2010,George2017}. In this work, we focus our attention on the two-qubit subregister composed of Q0 and Q2. The choice of Q2 stems from its central position in the device matrix, which is essential for implementing two-qubit gates. \textcolor{black}{The characteristic electrodynamical parameters of Q0 and Q2 are reported in Tab.~\ref{Tab1}, including statistical values of coherence times, readout pulse duration, $\pi$-pulse amplitude and the qubit-qubit coupling strength $J$, calculated following the procedure in App.~\ref{App3}.}
\begin{table*}[t]
\caption{\textcolor{black}{Summary of the parameters of the two-qubit subregister of the sQPU: the qubit frequency transition $\nu_{01}$ at the sweet spot (SS), the charging energy $E_c$, the statistical relaxation time $T_1$, the statistical Hahn-echo time $T_2^{Echo}$, the qubit-qubit coupling $J$ (App.~\ref{App3}), the readout pulse duration and the $\pi$-pulse amplitude.}}
\centering
\label{Tab1}
\begin{tabular}{cccc}
      \hline\noalign{\smallskip}
      &Q0 & &Q2 \\
      \hline\noalign{\vspace{1mm}}
      $\nu_{01}$ at SS [$GHz$] & $4.5546 \pm 0.0003$ & &$5.6503 \pm 0.0003$ \\
      {\vspace{1mm}}
      $E_c$ [$MHz$] &$340\pm2$ &&$274 \pm 2$ \\
      {\vspace{1mm}}
      $T_{1}$ [$\mu s$]&$24 \pm 5$ & &$8 \pm 1$ \\
      {\smallskip}
      $T_{2}^{Echo}$ [$\mu s$] & $10 \pm 3$ & &$6 \pm 1$ \\
      {\smallskip}
      J [$MHz$] &&$ 12\pm2 $& \\
      {\smallskip}
      Readout pulse duration [$ns$] &&200 $\pm$ 4& \\
      {\smallskip}
      $\pi$-pulse amplitude [$V$] &$0.155\pm0.001$&&$0.153\pm0.001$
\end{tabular}
\end{table*}
\textcolor{black}{An example of coherence times $T_1$ and $T_2^{Echo}$ for Q0 and Q2 are also reported in Fig.~\ref{chip5qubit} (b) and (c), respectively}.
Details on the characterization of the full matrix \textcolor{black}{can be found} in Refs.~\cite{Ahmad2023,Ahmad2024} \textcolor{black}{and the Supplementary Material of Ref.~\cite{Ahmad2024}}.

In order to experimentally characterize the sQPU, and then implement single- and two-qubit gates, we have thermally anchored the device to the lowest temperature stage of a Triton 400 dilution refrigerator~\cite{Ahmad2023,Ahmad2024}, equipped with coaxial cryogenic RF lines for control, readout, and tunability of the qubit frequency. A microwave FPGA (Field Programmable Gate Array) control electronics has been employed. Details of the cryogenic and room temperature components of the experimental setup are reported in App.~\ref{App1}. \textcolor{black}{Since the coherence times of the qubits are of the order of tens of microseconds (Tab.~\ref{Tab1}), the drive and the readout pulse durations have been chosen to be on the order of tens and hundreds of nanoseconds, respectively, to neglect the impact of decay on the single-qubit fidelity. Specifically,} \textit{XY} gates have been implemented by using 20 $ns$-long microwave drive pulses with a DRAG (Derivative Reductive Adiabatic Gate) shape to reduce leakage and undesired phase rotations~\cite{Lucero2008,Motzoi2009,reed2013,Werninghaus2021,Babu2021} (see Fig.~\ref{chip5qubit} (a) for a schematic representation of DRAG drive pulses). To optimize the parameters of drive pulses, such as the amplitude, the frequency and the Motzoi parameter~\cite{reed2013}, we first performed Rabi oscillations experiments as a function of the drive amplitude, and then iteratively applied combinations of Motzoi calibration, Ramsey protocol and AllXY technique as calibration routine~\cite{Lucero2008,Motzoi2009,reed2013,Werninghaus2021,Babu2021}. A detailed discussion on the drive pulses calibration procedure followed here has been reported in Ref.~\cite{Ahmad2024}. Finally, for what concerns \textit{CZ} gates, we employed standard unipolar flux pulses, whose amplitude and duration have been initially calibrated through Chevron experiments. We further calibrate the gate by performing conditional oscillation experiments around this optimal point~\cite{Rol2019}. A detailed discussion of the calibration procedure is reported in App.~\ref{app2}.

\subsection{Readout technique \textcolor{black}{and analysis}}

\begin{figure}[t]
    \begin{center}
    \includegraphics[width=0.49\textwidth]{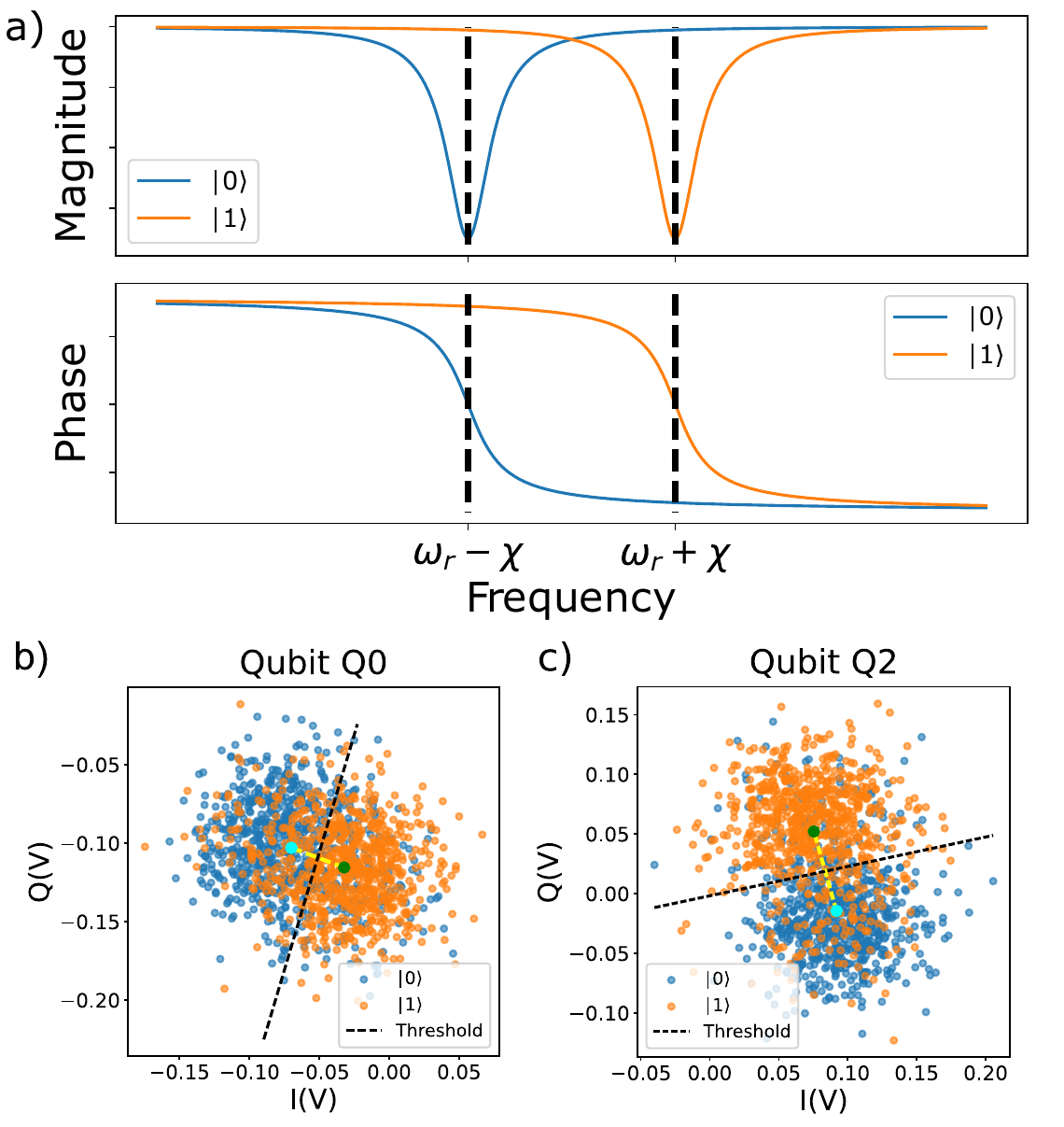}
    \end{center}
    \caption{In a), response of the readout resonator in terms of magnitude and phase when a qubit is prepared in the ground state (blue) and in the excited state (orange). The dashed lines indicate the resonator frequencies $\omega_r - \chi$, when the qubit is prepared in the ground state, and $\omega_r + \chi$ when it is prepared in the excited state, where $\omega_r$ and $\chi$ represent the resonance frequency of the isolated resonator and the dispersive shift, respectively~\cite{krantz2019}. In b) and c), single-shot readout calibration experiment for Q0 and Q2, respectively, where the blue and the orange scatter data correspond to the measurements of the qubit prepared in the ground and excited states for $760$ shots. \textcolor{black}{The cyan and green dots represent the centers of the blue and orange blobs, respectively. The dashed yellow line connects the centers of the two blobs, and the dashed black line represents the state-discrimination threshold.}}
    \label{thres}
\end{figure}

In order to analyze the quality of aggregates of single- and two-qubit gates, also known as quantum circuits, it is necessary to evaluate their output by reading out the qubit states on the computational basis. For single-qubit readout, the computational basis refers to the z-basis states, i.e. \{$\ket{0}$, $\ket{1}$\}. To map the voltage response of the superconducting resonator coupled with the qubit to a readout tone in the dispersive regime, we first perform a readout calibration. A schematic representation of the response of the readout resonator to a square pulse tone as a function of the frequency is reported in Fig.~\ref{thres} (a). 
In the dispersive regime, i.\,e. for sufficiently low power of the readout excitation signal, the resonant frequency of the resonator is shifted by $\chi$, according to the state of the qubit (blue for state $\ket{0}$ and orange for state $\ket{1}$). Therefore, by fixing the readout tone frequency to the one measured for the qubit in the ground state, variations in the voltage across the resonator identify changes in the qubit state. 

The readout calibration routine consists of preparing the qubit in each of the states of the computational basis repeatedly for a total number of shots. Here, we use $N_{shot}=760$ shots to ensure reasonable readout efficiency while remaining within the memory sampling limits of the electronics. Without any type of averaging, and for each state preparation, we measure the voltage response of the readout superconducting resonator (single-shot readout), obtaining a distribution of values. The voltage readout is a complex physical quantity, with a real and imaginary part. From now on, we will refer to these components as In-phase (I) and Quadrature-phase (Q), respectively. The experimental outcome is composed by two blobs in the IQ plane: one related to the measurement of the readout voltage when the qubit is prepared in $\ket{0}$, and the other when the qubit is prepared in $\ket{1}$ (in blue and orange in Fig.~\ref{thres}, respectively). 

\begin{figure*}[t]
    \begin{center}
\includegraphics[width=1\textwidth]{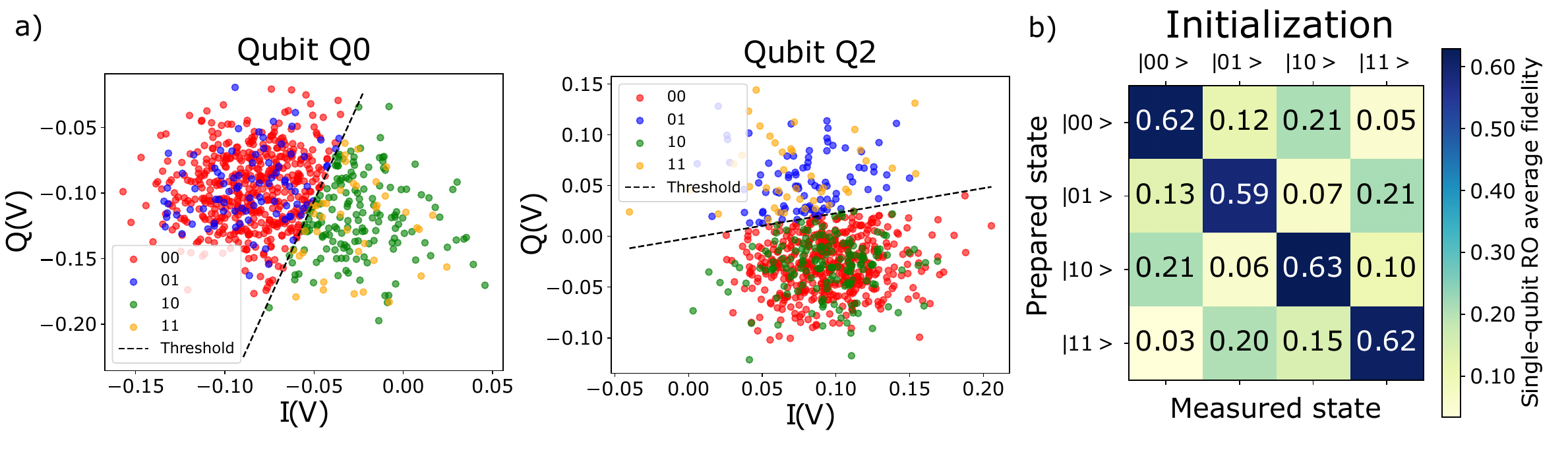}
    \end{center}
    \caption{In a), the measurement and thresholding technique for $\ket{00}$ state initialization is shown for Q0 and Q2. The red scatter data points corresponding to the simultaneous occurrance of $\ket{0}$ state readout for both qubits, i.e. $\ket{00}$. The blue dots correspond to measuring the $\ket{0}$ state on Q0 and the $\ket{1}$ state on Q2. The green ones correspond to the opposite situation, while the yellow dots correspond to measuring both qubits in the excited state. The black dashed line represents the threshold, defined as the midpoint of the line connecting the centres of the two blobs. The discrimination threshold is the same as in Fig.~\ref{thres}. In b) the probability matrix for two-qubit initialization. \textcolor{black}{The initialization experiment has been repeated 10 times, and the error, calculated as the semi dispersion, is around $1-2\%$.}}
    \label{init_2q}
\end{figure*}
To assign a binary value corresponding to the outcome of the qubit measurement, we \textcolor{black}{define} an assignment threshold, which best allows to distinguish between the prepared qubit states. \textcolor{black}{We first identify the coordinates of the midpoint of the blobs corresponding to $\ket{0}$ and $\ket{1}$ states (cyan and green dots in Fig.~\ref{thres} (b)-(c), respectively). The state discrimination threshold (dashed black line) is defined by the line perpendicular to the dashed yellow line in Fig.~\ref{thres} (b)-(c), which connects the two blobs midpoints~\cite{Ahmad2024}. As one can notice, not all the measured voltage values falls deterministically above or below the threshold. This is mainly related to two limitations: the signal-to-noise ratio (SNR) of the readout technique and the coherence performances of the qubit~\cite{Reed2010,krantz2019,Degraaf2020,Ahmad2024}. The readout blobs indirectly measure stochastic noise fluctuations due to qubit relaxation or dephasing phenomena. The former can be optimized by opportunely calibrating the readout tone power and duration, as well as the readout frequency~\cite{Mallet2009,Heinsoo2018,krantz2019,Chen2023}, or by integrating into the experimental setup advanced cryogenic amplification stages able to improve the SNR (e.g. superconducting near-quantum noise limited amplifiers)~\cite{Aumentando2020,Riste2012,Schmitt2014,Walter2017}. The latter is an intrinsic limitation of the qubit, which plays a fundamental role in NISQ devices. Therefore, a practical way to discriminate between different states of the qubit is to count how often the readout voltage falls above or below the set threshold, once the blobs separation in the IQ plane and the readout SNR have been optimized. This allows the definition of the count vectors $n_{ij}$, from which one can calculate the readout probability $p_{ij}=n_{ij}/N_{shot}$ to measure an assigned state while preparing the qubit on the computational basis. Here, the first index labels the measured state, while the second labels the prepared state.  
\textcolor{black}{To clarify, let's consider Fig.~\ref{thres} (b), and the readout calibration for Q0. For the preparation of the ground state (blue points), each shot is represented by a point in the IQ plane $\left(I_m,Q_m\right)$, where $m=\{1,\dots ,N_{shot}\}$. The discrimination threshold is represented by the linear function $I(Q)=A+BQ$, where $A$ is the intercept and $B$ is the slope. If $I_m > I(Q_m)$, we add a count to $n_{00}$; otherwise, we add a count to $n_{10}$. In contrast, for the preparation of the excited state (orange points), if $I_m > I(Q_m)$, we add a count to $n_{11}$; otherwise, we add a count to $n_{01}$. Finally, we obtain the readout probabilities by normalizing the count vectors to the total number of shots. The same procedure has been applied to Q2.}
This \textcolor{black}{analysis approach} is quite general and gives a fair estimation of the readout fidelity in isolated superconducting qubits~\cite{Mallet2009,Heinsoo2018,krantz2019,Chen2023}. However, when increasing the number of qubits, and specifically when dealing with two-qubit gates implementation, decoherence and entanglement may play a fundamental role, and it requires special care in the readout \textcolor{black}{analysis}.}

To stress this point, in this work we applied both single- and two-qubit gates, as well as a combination of the formers, on a two-qubit register, where the computational basis reads as $\{\ket{00}, \ket{01}, \ket{10}, \ket{11}\}$. When applying single-qubit gates, we set the qubits in their flux sweet spot (SS). Having Q0 a lower frequency than Q2, and specifically at least \textcolor{black}{$1 GHz\gg J$ (App.~\ref{App3})} lower than Q2 frequency~\cite{Ahmad2024}, we work within the single-qubit regime, meaning that the two qubits can be safely considered independent. Therefore, we can calculate the two-qubit readout count probability as the tensor product of the readout probability matrices of the single-qubit gates \textcolor{black}{calculated as above},
\begin{equation}
\hspace{2.5cm}
p_{ijkl}^{Q0,Q2}=p_{ij}^{Q0}\otimes p_{kl}^{Q2},
\label{eq10}
\end{equation}
where $\{i,j\}$ and $\{k,l\}$ label the single-qubit basis states for Q2 (control) and Q0 (target), respectively. The situation completely changes when considering two-qubit gates due to the entanglement effect. As mentioned earlier, two-qubit gates, and specifically the \textit{CZ} gate, require the qubits to be brought into resonance. This makes \textcolor{black}{Eq.\ref{eq10}} unsuitable.

We here detail on a method to \textcolor{black}{calculate the two-qubit readout probabilities},  which takes into account entanglement. The main difference from the previous technique lies in how to count the states. Since the system may show entanglement, it is not possible to calculate the probability of obtaining a two-qubit state as the tensor product of the coefficients of the single-qubit matrices. Indeed, one needs to include in the calculation of the two-qubit readout probability matrix the conditional probability. For simplicity, let's consider the following scenario. Let us prepare the system in the state $\ket{00}$, and perform a readout. We will have different probabilities of finding the system in the different basis states $\{\ket{00}$, $\ket{01}$, $\ket{10}$, $\ket{11}\}$. To define the count vector for $\ket{00}$ state, we count all instances where the two qubits simultaneously fall into the $\ket{0}$ state. \textcolor{black}{To clarify, let's consider Fig.~\ref{init_2q} (a) and (b), where we have highlighted with different colors the data points in Fig.~\ref{thres} (a) and (b). Specifically, if the conditions $I^{Q0}_m > I^{Q0}(Q^{Q0}_m)$ and $I^{Q2}_m < I^{Q2}(Q^{Q2}_m)$ are simultaneously met (red points), $p^{Q0,Q2}_{0000}$ increases. If Q0 is measured in the ground state, i. e., $I^{Q0}_m > I^{Q0}(Q^{Q0}_m)$, but simultaneously Q2 is instead measured in the excited state, i. e., $I^{Q2}_m > I^{Q2}(Q^{Q2}_m)$, $p^{Q0,Q2}_{0010}$ increases (blue points). In the same way, $p^{Q0,Q2}_{1000}$ will increase if $I^{Q0}_m < I^{Q0}(Q^{Q0}_m)$ and $I^{Q2}_m < I^{Q2}(Q^{Q2}_m)$ (green points), while $p^{Q0,Q2}_{1010}$ will increase if $I^{Q0}_m < I^{Q0}(Q^{Q0}_m)$ and $I^{Q2}_m > I^{Q2}(Q^{Q2}_m)$ (yellow points). A summary of this counting procedure is shown in Tab.~\ref{Tab_counts}.}

\begin{table}[h]
\caption{\textcolor{black}{Schematic representation for the counting of the states for the two-qubit register $\{Q0, Q2\}$.}}
\centering
\label{Tab_counts}
\renewcommand{\arraystretch}{1.5}
\begin{tabular}{c|cc}
      &$I^{Q0}_m > I^{Q0}(Q^{Q0}_m)$ & $I^{Q0}_m < I^{Q0}(Q^{Q0}_m)$ \\
      \hline
      $I^{Q2}_m < I^{Q2}(Q^{Q2}_m)$ &$\ket{00}$ &$\ket{10}$ \\
      $I^{Q2}_m > I^{Q2}(Q^{Q2}_m)$ &$\ket{01}$ &$\ket{11}$
\end{tabular}
\end{table}
\noindent
The readout probability matrix is shown in Fig.~\ref{init_2q} (b).
\section{Results}
\begin{figure*}[t]
    \begin{center}
    \includegraphics[width=1\textwidth]{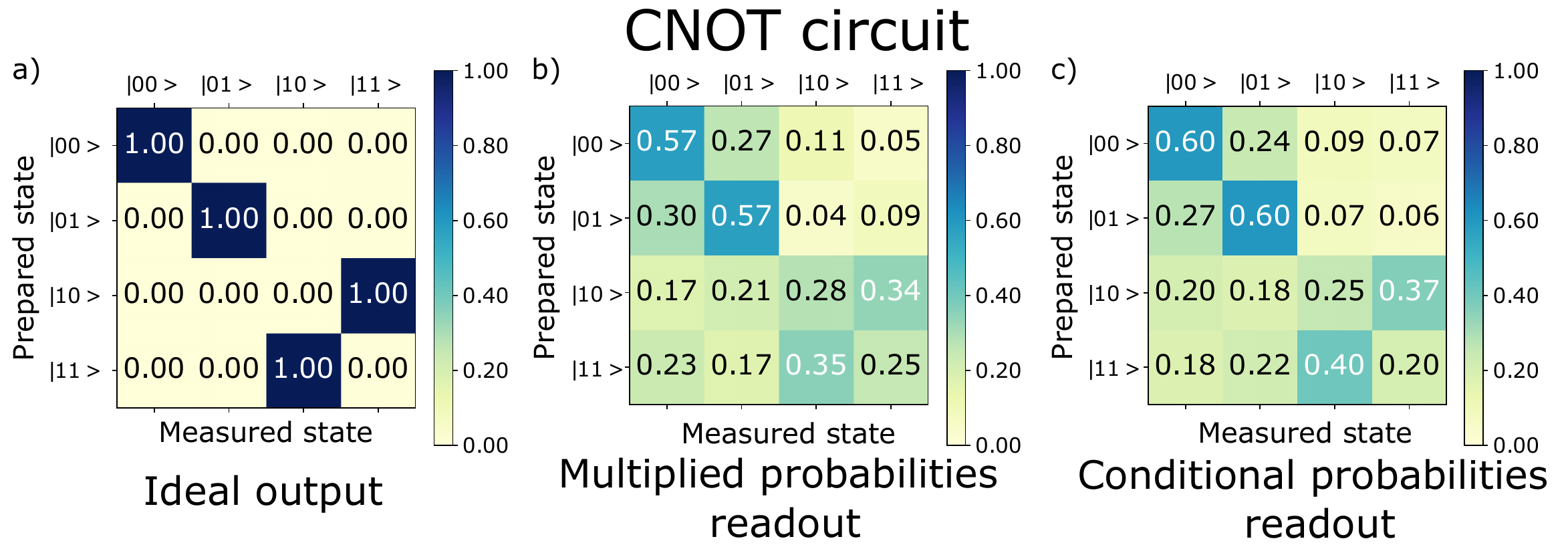}
    \end{center}
    \caption{\textcolor{black}{In a), the ideal probability matrix of the \textit{CNOT} quantum circuit over the two-qubit computational basis $\{\ket{control},\ket{target}\}$, where the control qubit is Q2 and the target is Q0. In b) and c), the measured probability matrix calculated using the tensor product of the single-qubit gate readout probabilities and the measured probability matrix calculated using the conditional readout paradigm, respectively. The color bar represents the readout probability. The experiment has been repeated 10 times, and the error, calculated as the semi dispersion, is ranging from $1\%$ to $6\%$.}}
    \label{CNOT_counts}
\end{figure*}
\begin{figure*}[b]
    \begin{center}
    \includegraphics[width=1\textwidth]{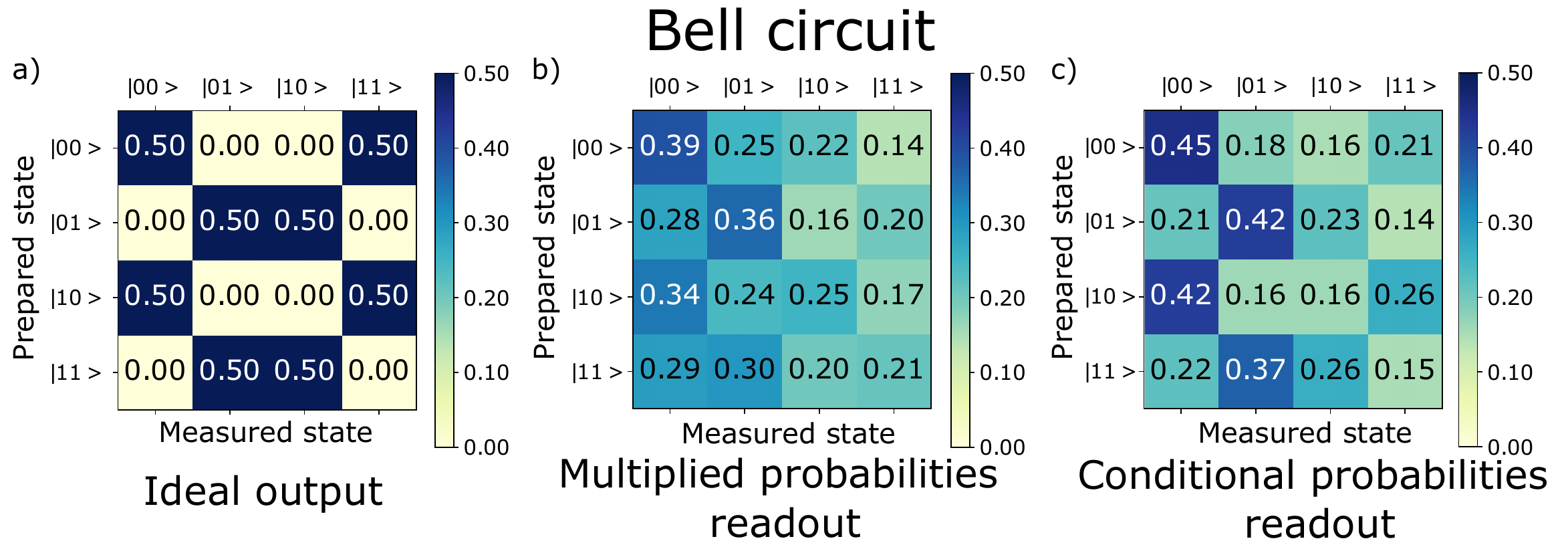}
    \end{center}
    \caption{\textcolor{black}{In a), the ideal probability matrix of the Bell quantum circuit over the two-qubit computational basis $\{\ket{control},\ket{target}\}$, where the control is Q2 and the target is Q0. In b) and c), the measured probability matrix calculated using the tensor product of the single-qubit gate readout probabilities and the measured probability matrix calculated using the conditional readout paradigm, respectively. The color bar represents the readout probability. The experiment has been repeated 5 times, and the error, calculated as the semi dispersion, is ranging from $1\%$ to $3\%$.}}
    \label{bell_prob}
\end{figure*}
From the initialization experiment and two-qubit state readout calibration in Sec.~\ref{Sec2}, the experimental probability matrix for two-qubit initialization (Fig~\ref{init_2q} (a)) shows that as long as Q2 is prepared in the ground state, we have a readout fidelity that is fairly in agreement with the results reported in the literature~\cite{Dasgupta2021}. Indeed, if we prepare the system in the state $\ket{00}$ or $\ket{01}$, the readout fidelity is above $60\%$. The lowest fidelities are reached when Q2 is prepared in the excited state. As discussed in Ref.~\cite{Ahmad2024}, Q2 exhibits lower coherence times than Q0, which may be accounted to the fact that it has a higher qubit frequency and the strongest connectivity with neighboring qubits. For this reason, Q0 is used as the target qubit in the implementation of the two-qubit quantum circuits tested in this work, including the Bell state generation, while Q2 is used as the control qubit.

\begin{figure*}[t]
    \begin{center}
    \includegraphics[width=1\textwidth]{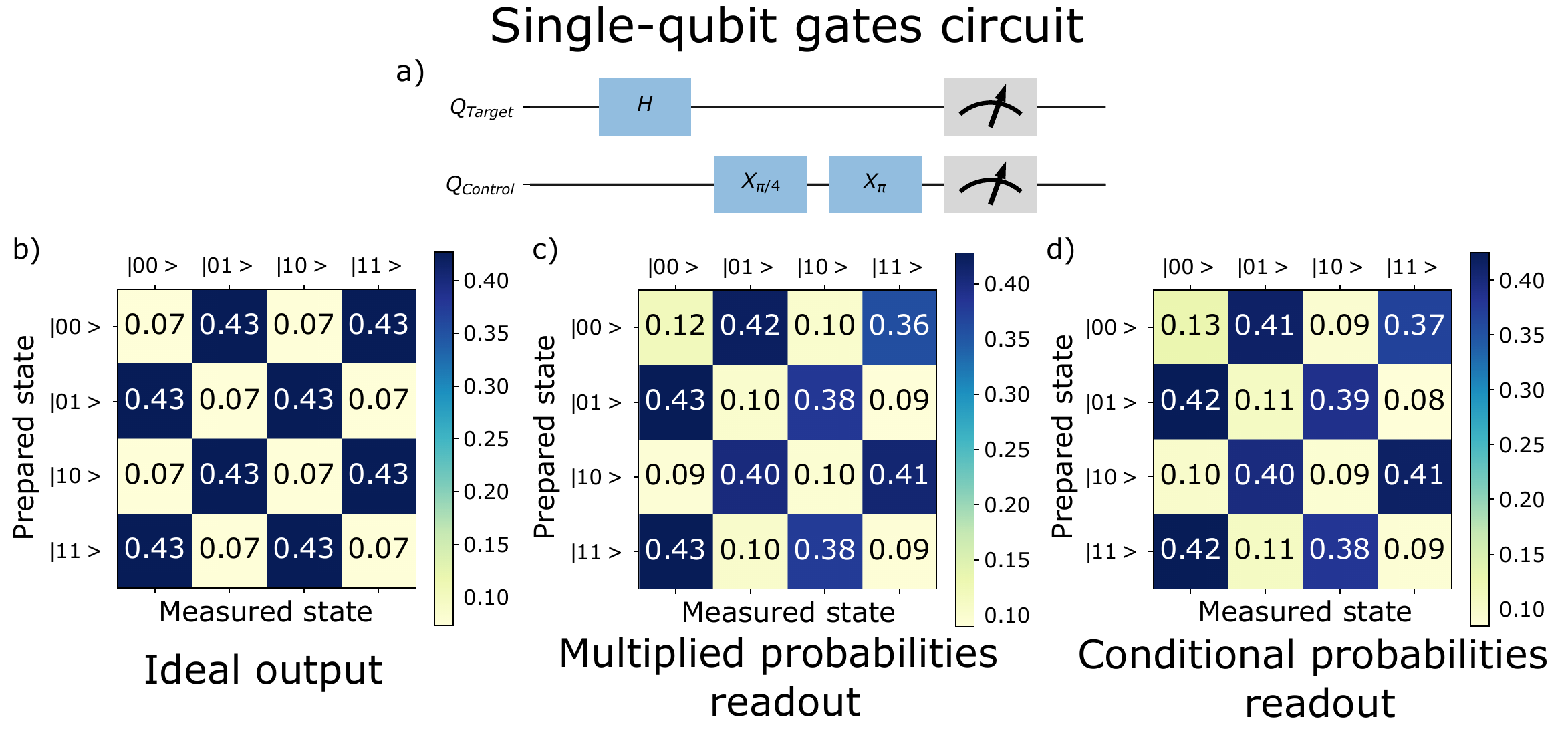}
    \end{center}
    \caption{\textcolor{black}{In a), the pulse scheme of the single-qubit gate random circuit. In b) and c), the ideal readout probability matrix and the measured probability matrix calculated using the product of the single qubit probabilities. In d), the measured probability matrix calculated using the conditional readout analysis. The color bar represents the probability distribution. The computational basis is the same used for the \textit{CNOT} quantum circuit. The experiment has been repeated 5 times, and the error, calculated as the semi dispersion, is ranging from $1\%$ to $3\%$.}}
    \label{single_qubit}
\end{figure*}

Towards the Bell circuit implementation, we have first applied a \textit{CNOT} on the two qubits of the sub-register (Fig.~\ref{CNOT_counts}), and we have calculated the output readout probability matrix over the two-qubit computational basis. This has been done by using both the multiplied \textcolor{black}{readout} probabilities and conditional \textcolor{black}{readout} probabilities \textcolor{black}{paradigms}. The two outputs of this preliminary circuit are compared with the ideal one calculated analytically. The results obtained highlight the necessity of using the conditional probabilities readout \textcolor{black}{paradigm} to analyze circuits involving two-qubit gates. In Fig.~\ref{bell_prob}, we report the implementation of the Bell quantum circuit with the qubits prepared in the initial state $\ket{00}$. A random single-qubit gate quantum circuit involving Hadamard and single-qubit gates has been also implemented to assess the performances of the technique for both entangling and non-entangling gates. The pulse scheme and the output matrices with the two techniques, as well as the comparison with the ideal output, are shown in Fig.~\ref{single_qubit}.
Finally, to compare the efficiency between the two readout \textcolor{black}{paradigms}, we use the Hellinger fidelity, a metric able to quantify the similarity between two classical discrete probability distributions~\cite{Hellinger1909}. \textcolor{black}{It} is defined as~\cite{Aktar2022}:
\begin{equation}
\hspace{2.2cm}
    H(p,q) = \Bigg[\sum_{i=1}^{n} \sqrt{p_i q_i}\Bigg]^{\textcolor{black}{2}},
\end{equation}
where \textcolor{black}{$p_i$ and $q_i$ are two discrete probability distributions, and $n$ represents the number of possible states of these distributions. The Hellinger fidelity can also be written in terms of the Hellinger distance $D_{H}$} as~\cite{Aktar2022}: 
\begin{equation}
\hspace{2.2cm}
    H(p,q) = (1-\textcolor{black}{D_{H}}^2)^2,
    \label{hell_fid}
\end{equation}
with $\textcolor{black}{D_{H}}$ defined as:
\begin{equation}
\hspace{1.5cm}
    \textcolor{black}{D_{H}} = \frac{1}{\sqrt{2}} \sqrt{\sum_{i=1}^{n}(\sqrt{p_i}-\sqrt{q_i})^2}.
\end{equation}
Specifically, the Hellinger fidelity is calculated in terms of the experimental probability readout state vector output of a quantum circuit for the device and the expected theoretical probability readout state vector for the same quantum circuit. We have calculated the Hellinger fidelity for all the implemented circuits using both the readout \textcolor{black}{analysis paradigms}. The results are shown in Tab.~\ref{Tab2}. 

\begin{table*}[b]
\caption{The \textcolor{black}{average} Hellinger fidelity for Single-qubit gate random circuit involving a Hadamard, an $\textcolor{black}{R_x(\pi/4)}$ and an $\textcolor{black}{R_x(\pi)}$ (Fig.~\ref{single_qubit}), the \textit{CNOT} and Bell quantum circuits calculated using both the discussed readout \textcolor{black}{paradigms}. The first label of the states refers to the control and the second to the target. \textcolor{black}{The error was calculated as the semi dispersion over 10 experiments for the single-qubit gate circuit and the CNOT, and 5 experiments for the Bell circuit.}}
\centering
\label{Tab2}
\begin{tabular}{lllllllll}
      \hline\noalign{\smallskip}
      &\multicolumn{8}{c}{Hellinger fidelity \textcolor{black}{[\%]}} \\
      \cline{2-9}\noalign{\smallskip}
      \multicolumn{1}{c}{Circuit} &\multicolumn{4}{l}{Multiplied probabilities} & \multicolumn{4}{l}{Conditional probabilities} \\
      \cline{2-9}\noalign{\smallskip}
      &$\ket{00}$ & $\ket{01}$ & $\ket{10}$ & $\ket{11}$ & $\ket{00}$ & $\ket{01}$ & $\ket{10}$ & $\ket{11}$ \\
      \hline\noalign{\smallskip}
      Single-qubit gate &$\textcolor{black}{99.2 \pm 0.5}$ &$\textcolor{black}{99.7 \pm 0.2}$ &$\textcolor{black}{99.2 \pm 0.6}$ &$\textcolor{black}{99.6 \pm 0.4}$ &$\textcolor{black}{99.2 \pm 0.5}$ &$\textcolor{black}{99.7 \pm 0.2}$ &$\textcolor{black}{99.2 \pm 0.7}$ &$\textcolor{black}{99.6 \pm 0.4}$ \\
      CNOT & $\textcolor{black}{59 \pm 3}$& $\textcolor{black}{55 \pm 6}$& $\textcolor{black}{33 \pm 5}$& $\textcolor{black}{37 \pm 3}$& $\textcolor{black}{61 \pm 3}$& $\textcolor{black}{57 \pm 6}$& $\textcolor{black}{38 \pm 6}$& $\textcolor{black}{42 \pm 2}$ \\
      Bell &$\textcolor{black}{49.9 \pm 0.1}$ &$\textcolor{black}{49.6 \pm 0.1}$ &$\textcolor{black}{50.0 \pm 0.1}$ &$\textcolor{black}{49.6 \pm 0.3}$ &$\textcolor{black}{63 \pm 1}$ &$\textcolor{black}{63.1 \pm 0.7}$ &$\textcolor{black}{64 \pm 3}$ &$\textcolor{black}{63 \pm 1}$
\end{tabular}
\end{table*}
\section{Discussion and concluding remarks}

The comparison between the experimental and the ideal readout probability matrices allows us to evaluate the conditional readout. We have calculated the Hellinger fidelity for all the implemented circuits using both the readout \textcolor{black}{paradigms} and underline the importance of implementing conditional readout for the output of circuits involving entangled gates. 

As shown in Fig.~\ref{single_qubit}, the experimental output of a random quantum circuit involving only single-qubit gates, taken here as a reference, has a Hellinger fidelity close to the state-of-the-art of $\sim99\%$. According to literature, typical values of Hellinger distance range between $0.1$ and $0.4$, which leads to a Hellinger fidelity between $\sim 70\%$ and $\sim 99\%$~\cite{Vaidman1999,Dasgupta2021,Schulze2021,wadhia2023}. As expected, the comparison between the matrices in Fig.~\ref{single_qubit} (a) and (b), and the Hellinger fidelities in Tab.~\ref{Tab2} obtained with the two readout \textcolor{black}{paradigms}, confirm that there is no sensitive improvement by using the conditional probabilities readout for single-qubit gate circuit. Indeed, as mentioned in Sec.~\ref{Sec2}, the single-qubit gates were applied in the approximation of isolated qubits, i.\,e. the output states of the circuit can be considered as separable. 

On the other hand, as quantum circuits grow in complexity with an increasing number of qubits involved, specifically in the case of entangling gates like the \textit{CNOT} in Fig.~\ref{CNOT_counts}, the output quality is affected. We account for the lower Hellinger fidelities observed for two-qubit gate circuits compared with the single-qubit quantum circuit to the fidelity of the \textit{CNOT} gate itself~\cite{Ahmad2023}. Specifically, we have used unipolar \textit{CZ} flux pulses. It has been demonstrated that unipolar flux pulses are less efficient than other variants of the same gate, which uses more complicated flux pulse shaping~\cite{Rol2019,Negirneac2021}. Nevertheless, we can still notice that a fundamental role is also played by the readout \textcolor{black}{analysis}. In the case of the \textit{CNOT} quantum circuit, by taking explicitly into consideration the conditional probability in the two-qubit basis states discrimination procedure, we can improve the Hellinger fidelity by approximately $3\%$ and $5\%$, when the control qubit is prepared in the ground state and the excited state, respectively (Tab.~\ref{Tab2}). The need to apply the conditional probabilities readout is even more noticeable in the case of the Bell circuit in Fig.~\ref{bell_prob}. Just by using the single-qubit probability tensor product, the circuit output returns incorrect results in the majority of cases, since the Hellinger fidelity drops below $50\%$~\cite{wadhia2023}. Using the \textcolor{black}{conditional readout probability analysis}, instead, the Hellinger fidelity improves from $13\%$ to $17\%$, reaching fidelity values of $67\%$, well in line with typical values reported in the literature~\cite{Dasgupta2021}, despite the numerous challenges associated with the implementation of low-error and high-fidelity two-qubit gates. 

In conclusion, we have demonstrated that as soon as the complexity of quantum circuits increases, and specifically when dealing with maximally entangled states, the efficiency of quantum algorithms and the quality of their output, not only depend on the quality of the quantum gates implemented, but sensitively depends on the readout \textcolor{black}{analysis} used. Taking into account conditional readout \textcolor{black}{probabilities paradigm} while reading out the state of two or more qubits in their computational basis \textcolor{black}{is extremely useful} to correctly assess the output of quantum algorithms in the presence of entanglement. These results aim to provide a guide for those who want to implement quantum algorithms on NISQ processors for the first time, indicating in detail the necessary steps to benchmark the readout \textcolor{black}{probabilities depending} on the quantum circuit under test. 

\appendix

\section{Cryogenic and room-temperature setup}
\label{App1}
To characterize and analyze superconducting quantum devices it is fundamental to achieve the superconducting regime and minimize the effects of thermal noise during the measurements. Specifically for superconducting qubits, if they are sufficiently below the temperature $T = \hbar \omega_{01}/k_B$, transitions between the two computational levels $\ket{0}$ and $\ket{1}$ due to thermal fluctuations can be safely neglected~\cite{Krinner2019}. 
The dilution fridge employed in this work is the dry Triton 400 of the Oxford
Instruments, which is capable of reaching temperatures of about 10 $mK$. 

\begin{figure}[h!]
    \begin{center}
    \includegraphics[width=0.48\textwidth]{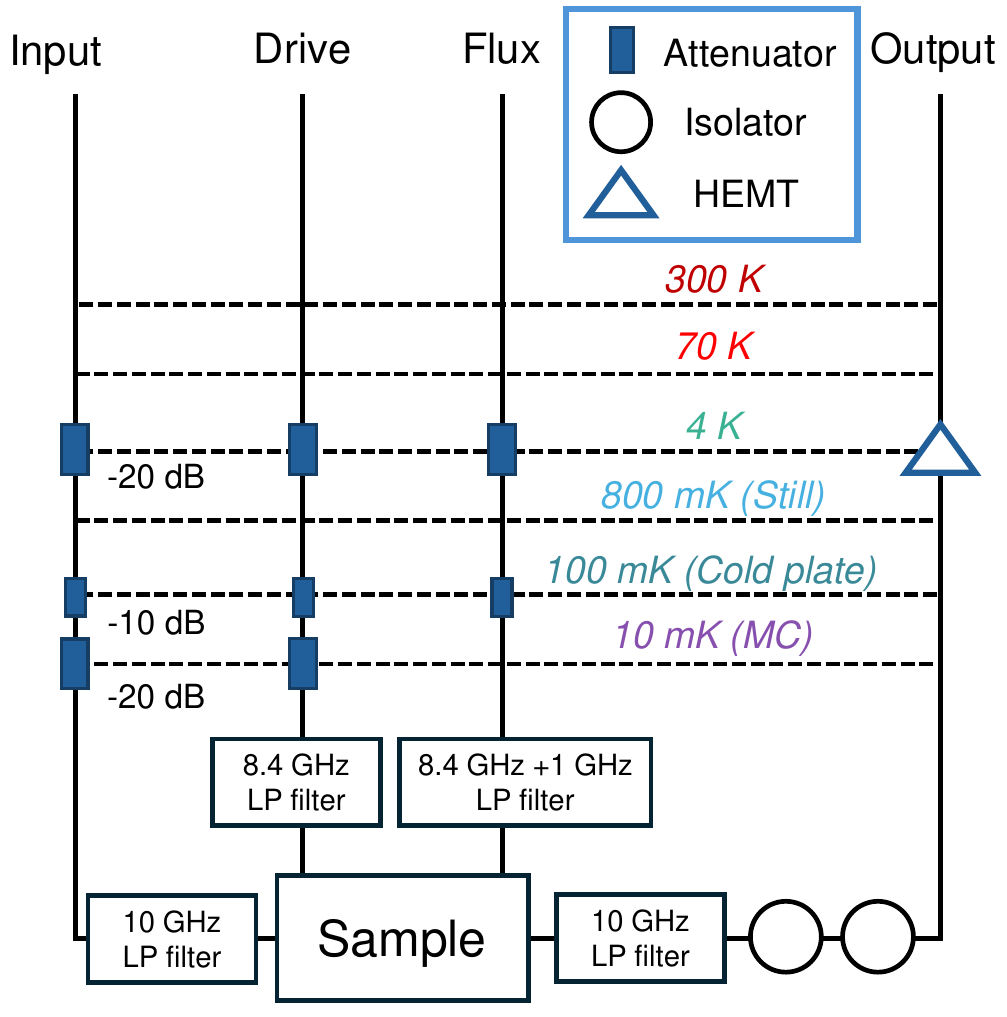}
    \end{center}
    \caption{Cryogenic setup scheme, including the attenuation scheme for the input, drive and flux lines. On the output line, there are two isolators and an HEMT amplifier. Each line has a low-pass filter.}
    \label{setupcryo}
\end{figure}

To characterize superconducting qubits, coaxial cables are used, which allow
operation within the microwave frequency range (DC - 18 $GHz$, nominally). The cryostat features four types of
lines: input and output lines for the readout, drive lines for qubit control, and flux lines
for frequency tuning with an external magnetic flux. Specifically, on the feedline and drive lines, there is an overall attenuation of -50 $dB$ and a low-pass filter with a cutoff of about 10 $GHz$ and 8.4 $GHz$, respectively. On the flux lines, there are -30 $dB$ of attenuation and two low-pass filters of 8.4 $GHz$ and 1 $GHz$. Finally, on the output lines, there is a 10 $GHz$ low-pass filter and two isolators. Signals from the output towards the sample and reflections at the input port are attenuated nominally by a total of 40 $dB$. Since the output signals of the qubits are single-photon signals, amplifiers are required. In our system, there are two amplification stages. There is a High Electron Mobility Transistor (HEMT) with nominal amplification of 40 $dB$ on the 4 $K$ plate, and three amplifiers at room temperature with a nominal 16 $dB$ amplification each. The cryogenic setup is shown in figure \ref{setupcryo}.
 \\
To implement algorithms at the pulse level, we have used a Qblox room-temperature microwave modular electronics fully interfaceable with Python, thanks to the open-source package Quantify~\cite{Quantify}. 
The modules used for the measurements of this work are:
\begin{itemize}
    \item Qubit Readout Module RF (QRM-RF), which allows output and input signals up to 18.5 $GHz$;
    \item Qubit Control Module RF (QCM-RF), which allows output signals up to 18.5 $GHz$;
    \item Qubit Control Module (QCM) for flux pulses, which generates output signals up to 400 $MHz$ and offsets to set the operation idle frequency of the investigated qubits.
\end{itemize}

\section{\textcolor{black}{Characterization of Qubit–Qubit Coupling}}
\label{App3}
\begin{figure}[h]
    \begin{center}
    \includegraphics[width=0.48\textwidth]{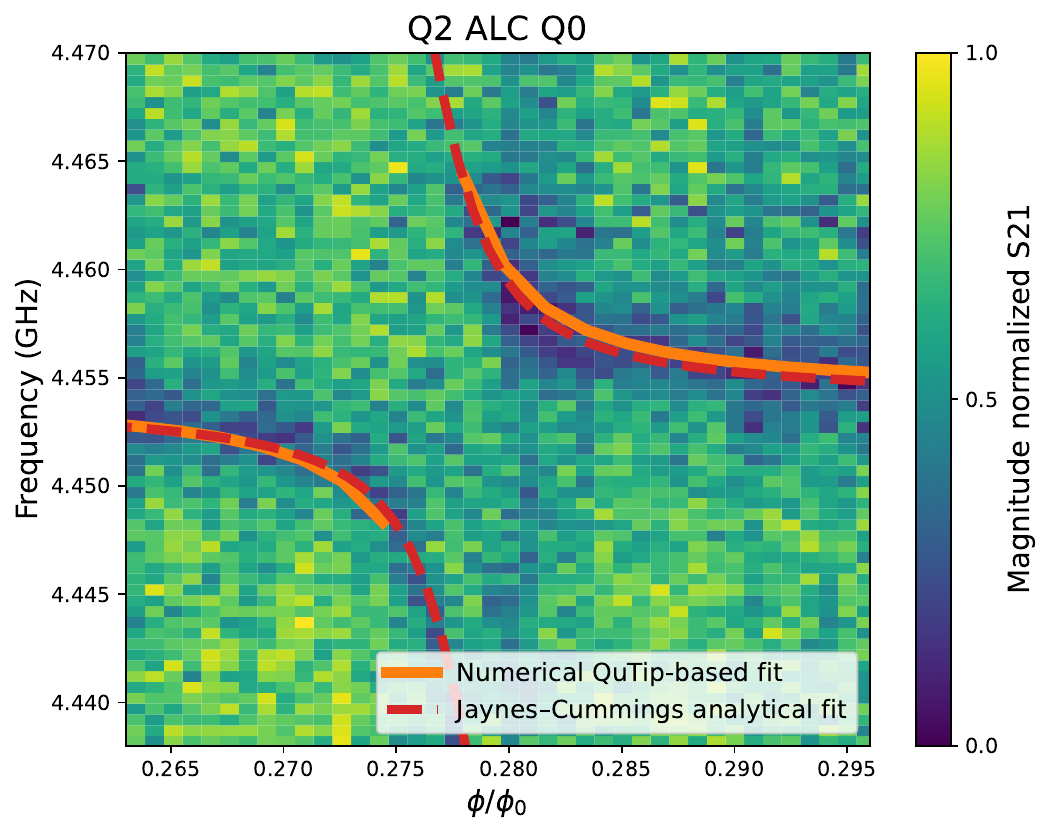}
    \end{center}
    \caption{\textcolor{black}{ALC for Q0 and Q2 between the $\ket{01}$ and $\ket{10}$ levels: on the x-axis the applied flux on Q2, on the y-axis the qubit frequency of Q0 and the color scale is the normalized voltage measured on the readout resonator 0. The red line represents the fit of the ALC using the Eq.~\ref{JC_diag_fit}, while the orange one is the numerical QuTip-based fit.}}
    \label{JC_q2q0}
\end{figure}

\textcolor{black}{The two-qubit register, coupled by means of a high-frequency bus resonator, is modeled by the complete Hamiltonian:}
\begin{equation}
\begin{split}
\hspace{0.6cm}
\textcolor{black}{
H = \sum_{i=1,2} \left( \nu_i\, b_i^\dagger b_i - \frac{E_{C_i}}{2}\, b_i^\dagger b_i^\dagger b_i b_i \right) + \nu_r\, a_r^\dagger a_r \nonumber +}\\
\textcolor{black}{
+ g_{1r}\left( b_1^\dagger a_r + b_1\, a_r^\dagger \right) \nonumber + g_{2r}\left( b_2^\dagger a_r + b_2\, a_r^\dagger \right),}
\end{split}    
\end{equation}
\textcolor{black}{
where the operators $b_i$ ($b_i^\dagger$) denote the annihilation (creation) operators for the transmon qubits (with \(i=1,2\)) and \(a_r\) (\(a_r^\dagger\)) are the annihilation (creation) operators for the resonator mode. Here, \(\nu_i\, b_i^\dagger b_i\) represents the bare energy of qubit \(i\), with the anharmonicity introduced by \(-\frac{E_{C_i}}{2}\, b_i^\dagger b_i^\dagger b_i b_i\), while \(\nu_r\, a_r^\dagger a_r\) accounts for the energy of the single resonator mode. The interaction between each qubit and the resonator is described by the terms \(g_{ir}(b_i^\dagger a_r + b_i\, a_r^\dagger)\). Operating in the dispersive regime ($|\Delta_i| = |\nu_i-\nu_r| \gg g_{ir}$), the resonator is only virtually populated, which allows the reduction of the full Hamiltonian to an effective Jaynes--Cummings model. In this limit, the effective qubit--qubit interaction is given by~\cite{Blais2007,Filipp2011}:}
\begin{equation}
\label{JC_diag_fit}
\hspace{0.4cm} \textcolor{black}{f_{01}(\nu_1, \nu_2, J)=\frac{(\nu_1 + \nu_2) \pm \sqrt{(\nu_1 - \nu_2)^2+4J^2}}{2}\hspace{2mm}_.}
\end{equation}
\textcolor{black}{
To estimate the effective coupling $J$ between Q0 and Q2, we performed an avoided level crossing (ALC) measurement, shown in Fig.~\ref{JC_q2q0}.}
\textcolor{black}{
In this experiment, spectroscopy was performed on Q0 while sweeping the flux on Q2, revealing the spectrum of the dressed eigenstates $\ket{01}$ and $\ket{10}$ at $\phi/\phi_0 \approx 0.278 $ flux quanta of Q2. Q0 was maintained at a fixed flux bias of $\phi/\phi_0 \approx 0.086$, and a drive tone with $86 dBm$ attenuation was applied to induce the $\ket{0}\rightarrow\ket{1}$ transition required for the measurement.
Fitting the experimental data with this model yields an effective coupling of \(J = (12 \pm 2)\,\mathrm{MHz}\), while numerical simulation of the eigenvalues of the complete Hamiltonian (via a QuTip-based approach~\cite{QuTip}) gives \(J = (10 \pm 2)\,\mathrm{MHz}\). Although the uncertainty of \(\pm 2\,\mathrm{MHz}\) (approximately 15–20\% of the nominal value) may appear large, it reflects the limited spectral resolution and noise of the measurement. The agreement between these two independent evaluations, within these error margins, confirms the robustness of the effective coupling model in capturing the non-perturbative dynamics of the coupled qubit system.}

\section{Two-qubit gates \textit{CZ} calibration}
\label{app2}
\begin{figure*}[t!]
    \begin{center}
    \includegraphics[width=1\textwidth]{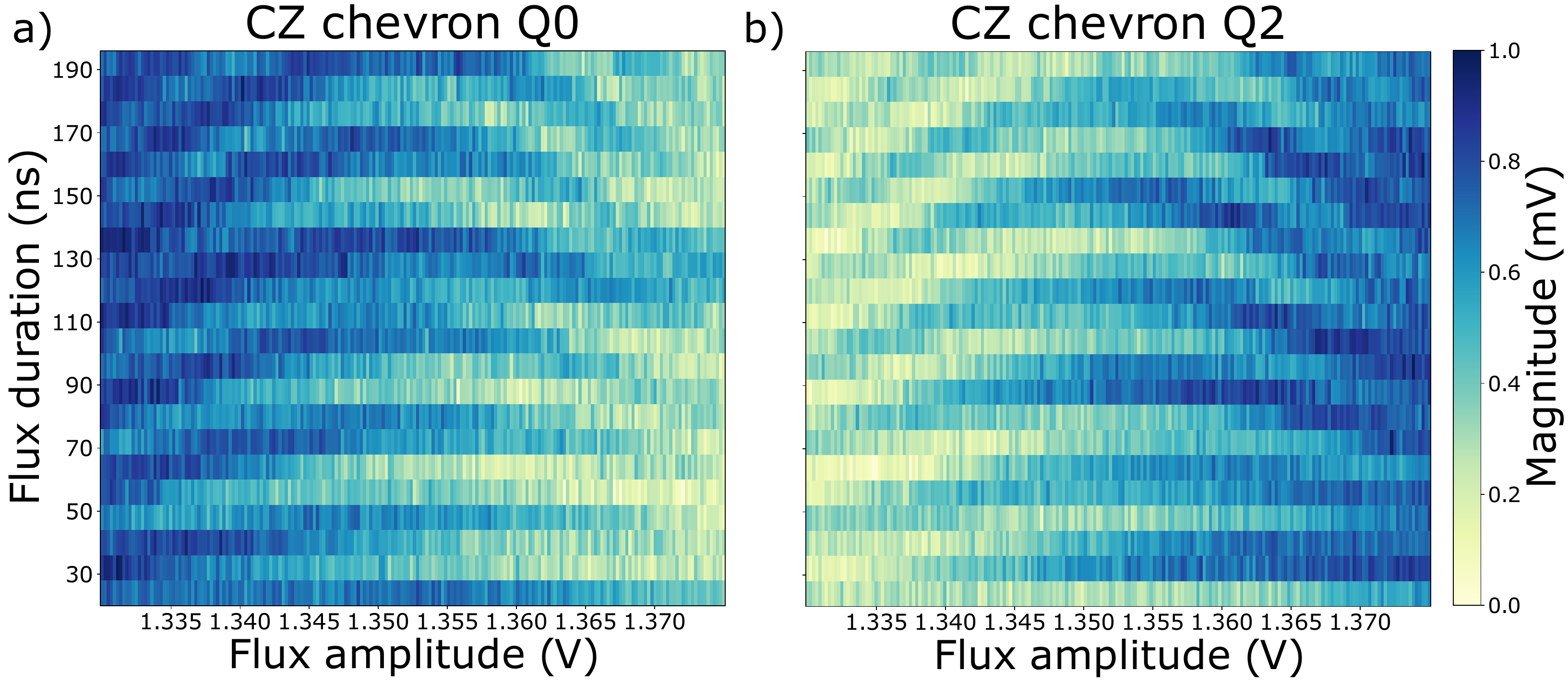}
    \end{center}
    \caption{Chevron plots for the \textit{CZ} experiment. On the y-axis the duration of the flux pulse on Q2, on the x-axis the magnitude of the flux amplitude and the color scale is the normalized magnitude of the readout resonators of a) Q0 and b) Q2. We perform a simultaneous measurement of both qubits, showing the characteristic coherent exchange of energy between the two coupled qubits.}
    \label{chev_CZ}
\end{figure*}

\begin{figure}[h!]
    \begin{center}
    \includegraphics[width=0.47\textwidth]{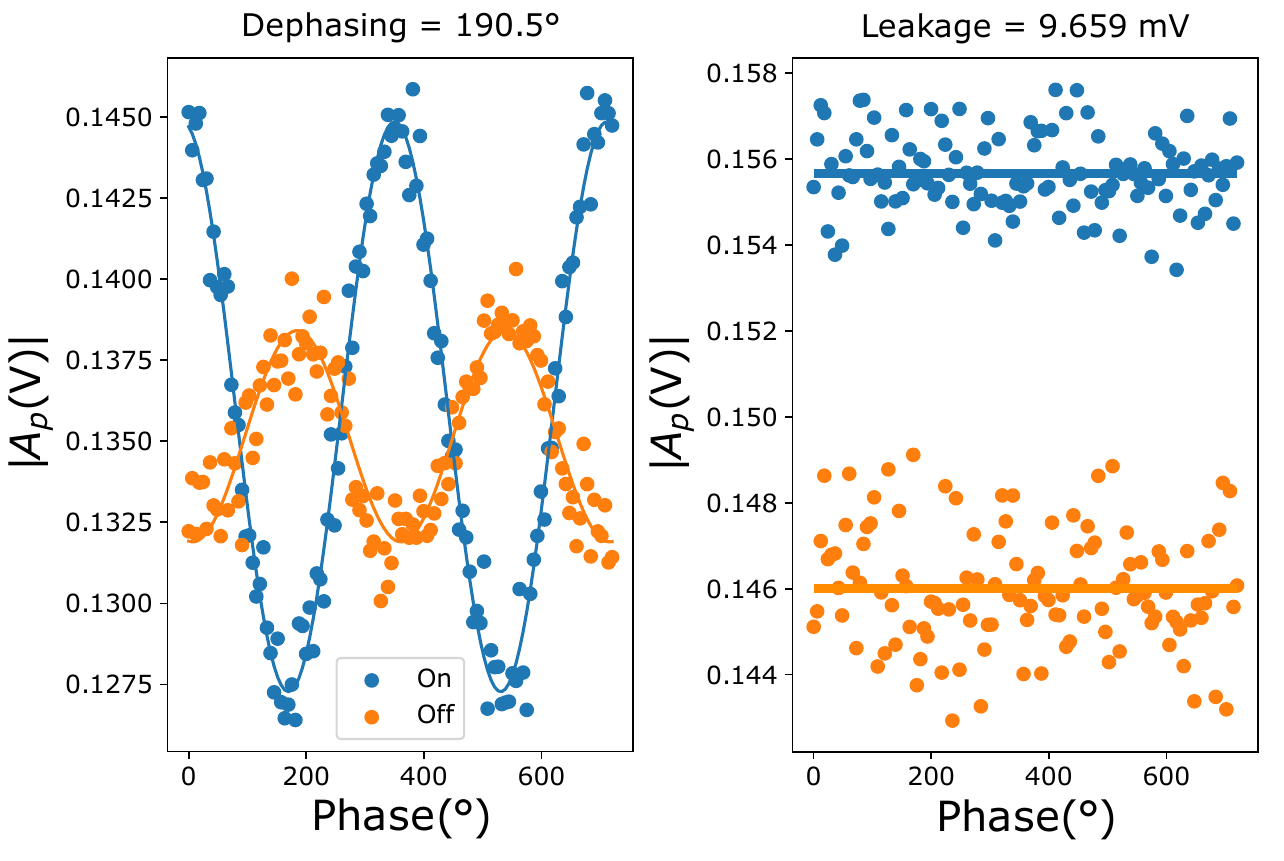}
    \end{center}
    \caption{The measured voltage as a function of the phase on the Q0 on the left and on Q2 on the right. The blue lines represent the measured values for the On variant, while the orange lines the measured values for the Off variant.}
    \label{cond_oscill_meas}
\end{figure}

The \textit{CZ} experiment requires to excite both the qubits so that the system is in $\ket{11}$. We fix the Q0 in its flux sweetspot and we change the flux pulse amplitude and duration on Q2. The results of this experiment measuring the Q0 and Q2 are shown in panels a) and b) of figure \ref{chev_CZ}. \\
The measurements are performed simultaneously on both qubits by sending a two-tone signal into the feedline. The first tone is tuned to resonate with resonator 0, and the second tone with resonator 2.
We extrapolate the oscillations at a fixed flux voltage pulse amplitude and use the Rabi oscillations approach to estimate the duration of the \textit{CZ} gate. The oscillations between the $\ket{11}$ and $\ket{02}$ states happen approximately in $(16 \pm 4) ns$. The oscillations shown in figure \ref{chev_CZ} are corrected from distortions, using a hardware solution called \textit{Cryoscope}~\cite{Rol2020}, which employs a series of filters.
The distortions can be due to many factors. A very common problem is the shape of the signal that we use for the flux biasing. The pulse used to implement the \textit{CZ} gate is a unipolar pulse~\cite{Rol2019}, which has a finite rise time. Moreover, the electronics generates signals that are not exactly square pulses and this can cause deviations in the measured response.
This behavior becomes particularly important when the \textit{CZ} duration is short. \\
Finally, to optimize the \textit{CZ} pulse parameters, we perform a conditional oscillation experiment~\cite{Rol2019}. It consists of two variants of the same experiment. For the Off variant, the target qubit, i.e. Q0, is prepared on the equator by a $\pi/2$-pulse, while the control qubit is left in the ground state. Then the \textit{CZ} flux
pulse is applied, followed by another $\pi/2$-pulse. Finally, the states of both qubits are measured simultaneously. For the On variant, the pulses on the Q0 are the same as the Off variant, but on the control, i.e. Q2, is applied a $\pi$-pulse each time a $\pi/2$-pulse is applied on Q0. If the phase difference between the readout signals measured on the target in the two configurations is not 180 degrees, the parameters of the \textit{CZ} gate are iteratively changed until this value is achieved. At the same time, the measurement on the control allows us to estimate the leakage, i.e. the probability that a random computational state leaks out of the computational subspace. The optimal amplitude and duration parameters of the \textit{CZ} pulse are the ones for which the phase difference is as close as possible to 180 degrees, while at the same time minimizing the leakages. The result of the best measurement for the conditional oscillation experiment is shown in figure \ref{cond_oscill_meas}, corresponding to a phase difference $\theta_{2Q} \simeq 191^{\circ}$ and leakage $L \simeq 9.66 mV$.
 \\

\begin{acknowledgements}

The work was supported by the Pathfinder EIC 2023 project "FERROMON-Ferrotransmons and Ferrogatemons for Scalable Superconducting Quantum Computers", the project SuperLink - Superconducting quantum-classical
linked computing systems, call QuantERA2 ERANET COFUND, CUP
B53C22003320005, the PNRR MUR project PE0000023-NQSTI and the
PNRR MUR project CN$\_$00000013 -ICSC, and the Project PRIN 2022-
Advanced Control and Readout of Scalable Superconducting NISQ Architectures (SuperNISQ)-CUP E53D23001910006.
We thank SUPERQUMAP project (COST Action
CA21144), and the invaluable technical support provided by the electronics and software engineers at Qblox (Delft, Netherlands), and Orange
Quantum Systems (Delft, Netherlands).
\textcolor{black}{We thank QuantWare for the kind demo of a non guaranteed device.}
We thank Roberto Schiattarella for fruitful discussions.

\end{acknowledgements}

\section*{Competing Interests and Funding}
The authors have no competing interests to declare that are relevant to the content of this article.

D. Ma. and F. T. are Editors for the special collection "Superconducting nanodevices: quantum and classical materials for coherent manipulation”. F. T. is part of the Board of Associate Editors of Journal of Superconductivity and Novel Magnetism.

\section*{Data availability statement}
Due to confidentiality agreements, the diagnostic data on the superconducting device supporting this study, so as the sample, can only be made available to personnel subject to a non-disclosure agreement. All other data can be made available by the authors upon reasonable request.

\bibliographystyle{ieeetr}
\bibliography{bibliography}

\end{document}